\documentclass[12pt,preprint,letterpaper]{aastex}
\newcommand{\fig}[1]{Figure~\ref{#1}}
\newcommand{\speed}[1]{#1 km~s${}^{-1}$}
\newcommand{\aspeed}[1]{$\sim\,$#1 km~s${}^{-1}$}

\begin{document}

\shorttitle{Evidence of An Unwinding Polar Jet} %

\shortauthors{Shen et al.}

\title{KINEMATICS AND FINE STRUCTURE OF AN UNWINDING POLAR JET OBSERVED BY {\sl SOLAR DYNAMIC OBSERVATORY}/ATMOSPHERIC IMAGING ASSEMBLY}

\author{Yuandeng Shen\altaffilmark{1,2,3}, Yu Liu\altaffilmark{1,3}, Jiangtao Su\altaffilmark{3,4}, and Ahmed Ibrahim\altaffilmark{5}}

\altaffiltext{1}{National Astronomical Observatories/Yunnan Observatory, Chinese Academy of Sciences, Kunming 650011, China; ydshen@ynao.ac.cn}
\altaffiltext{2}{Graduate University of Chinese Academy of Sciences, Beijing 100049, China}
\altaffiltext{3}{Key Laboratory of Solar Activity, National Astronomical Observatories, Chinese Academy of Science, Beijing 100012, China}
\altaffiltext{4}{National Astronomical Observatories, Chinese Academy of Sciences, Beijing 100012, China}
\altaffiltext{5}{Physics and Astronomy Department, College of Science, King Saud University, P.O. Box 2455, Riyadh 11451, Saudi Arabia}

\begin{abstract}
We present an observational study of the kinematics and fine structure of an unwinding polar jet, with high temporal and spatial observations taken by the Atmospheric Imaging Assembly on board the {\sl Solar Dynamic Observatory} and the Solar Magnetic Activity Research Telescope. During the rising period, the shape of the jet resembled a cylinder with helical structures on the surface, while the mass of the jet was mainly distributed on the cylinder's shell. In the radial direction, the jet expanded successively at its western side and underwent three distinct phases: the gradually expanding phase, the fast expanding phase, and the steady phase. Each phase lasted for about 12 minutes. The angular speed of the unwinding motion of the jet and the twist transferred into the outer corona during the eruption are estimated to be $11.1 \times 10^{-3}$ rad s$^{-1}$ (period = 564 s) and 1.17--2.55 turns (or 2.34--5.1$\pi$), respectively. On the other hand, by calculating the azimuthal component of the magnetic field in the jet and comparing the free energy stored in the non-potential magnetic field with the jet's total energy, we find that the non-potential magnetic field in the jet is enough to supply the energy for the ejection. These new observational results strongly support the scenario that the jets are driven by the magnetic twist, which is stored in the twisted closed field of a small bipole, and released through magnetic reconnection between the bipole and its ambient open field.
\end{abstract}


\section{Introduction}
Coronal jets are heated plasmas along open or large-scale magnetic field lines. They often have a linear or slightly bent structure. In general, magnetic reconnection is regarded as the basic driving mechanism for them \citep[e.g.,][]{shib96,chen08}. The detailed observational characteristics could be found in \cite{shim96}. On the other hand, H$\alpha$ surges, which are cold plasma flows and observed as emission in H$\alpha$ or absorption at other wavelengths, are usually considered to be cool counterparts of coronal jets in the chromosphere owing to the close relationship between them in space and time \citep{liu04,jian07}. It should be noted that sometimes surges can trigger large-scale coronal mass ejections \citep{liu05,liu08}.

Jets or surges (hereafter, we will refer to both as ``jets'') with rotary motion have been reported in a lot of studies \citep[e.g.,][]{xu84,okte90,canf96,pike98,alex99,harr01,jibb04,pats08,liu09,moor10}. However, the exact physical causes remain unclear due to the limitation of our observations. \cite{xu84} proposed that the double-pole diffusion of plasmas, which is resulted from the sharp density gradient between the ejecting plasmas and their surrounding open field, could help to interpret the jet's rotary motion. Whereas, others explained such phenomena as the product of magnetic reconnection between a small bipole with twisted closed field and a large-scale background open field \citep{canf96}. \cite{jibb04} adopted such a scenario to explain the twist propagation from the twisted closed field of the bipole to the ambient open field. They found that the direction of the observed spin of jets is in agreement with the relaxation of the twist stored in the closed field of the bipole. A 2.5-dimensional simulation proposed by \cite{shib86} showed that the mass and twist stored in the twisted closed field of the bipole could be driven out into the ambient open field. In addition, they further pointed out that there should be a helical velocity field in the jet consisting of a hot core and a cool sheath. Three-dimensional simulations also indicated that the releasing of stored twist does produce massive, high-speed jets driven by nonlinear Alfv$\rm \acute{e}$n waves in the solar corona \citep{pari09,pari10}. However, before the launching of {\sl Solar Dynamic Observatory (SDO)}, very few observations could provide enough information to confirm the close relationship between the plasma motions and the non-potential magnetic field energy released for the jets.

In this paper, we study the fine structure and kinematics of an unwinding coronal jet occurring in the north polar coronal hole on 2010 August 21. In addition, the energy relationship between the jet and its non-potential magnetic field are investigated. Observations and instruments are described in Section 2, results are presented in Section 3, discussions and conclusions are summarized in Section 4.

\section{INSTRUMENTS AND OBSERVATIONS}
The observations used in this paper are taken by {\sl SDO}/Atmospheric Imaging Assembly (AIA; \cite{titl06}) and Solar Magnetic Activity Research Telescope (SMART; \cite{ueno04}) instruments. Full-disk EUV images supplied by the AIA telescope on board the recently launched {\sl SDO} satellite have a cadence of 12 s and a spatial resolution of $1\arcsec.2$. The SMART can simultaneously provide full-disk images in five channels: H$\alpha$ line center and four off-bands ($\pm$0.5, $\pm$0.8 \AA). The images are taken with 1 minute cadence and $1\arcsec.2$ spatial resolution. All the images are differentially rotated to a reference time (06:40:00 UT). Meanwhile, they are rotated such as north (east) is up (left).

\section{RESULTS}
The jet started at about 06:07 UT and ended around 07:08 UT. It ejected along a straight trajectory resembling a cylinder with clear helical structures on the surface. In addition, obvious unwinding motions around the jet's main axis were observed during the rising period.

\begin{figure}\epsscale{1}
\plotone{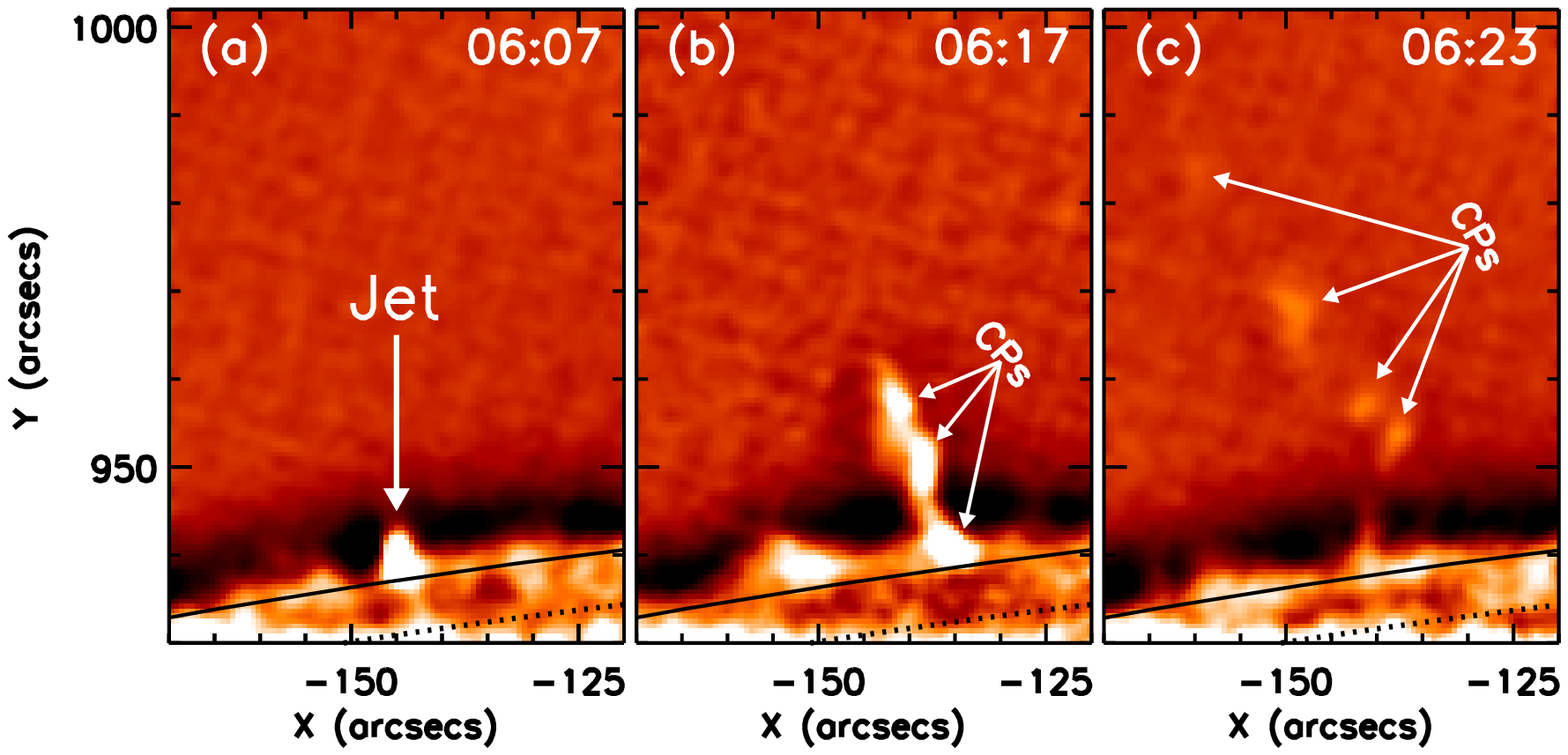}
\caption{Time sequence of SMART H$\alpha$ images. The jointed arrows indicate the condense points (CPs), while the white arrow in frame (a) marks the appearance of the jet. The black dotted curves in each frame are the latitude--longitude grid. The spacing between the curves is 20$^{\circ}$. Meanwhile, the solid black curve marks the solar disk limb (it is the same in the following figures). The field of view (FOV) is $50\arcsec \times 72\arcsec$ for each frame.}
\label{halpha}
\end{figure}

\fig{halpha} is a time sequence of H$\alpha$ images. At 06:07 UT, the jet appeared and displayed as a cuspate bright structure in H$\alpha$ observations near the northeast of the disk limb (see \fig{halpha}(a)). About 10 minutes later, it evolved into a collimated structure consisting of separate dense points (see the jointed white arrows in \fig{halpha}(b) and (c)). According to \cite{okte90}, these dense points represent the appearance of the rotary motion of the moving plasmas in a jet. The rising speed of the jet front is measured to be \speed{68} from the H$\alpha$ images, and the maximum height traced is up to 5 $\times 10^{4}$ km. It is worthwhile to note that various speeds discussed in this paper are the apparent speeds of propagating bright structures in the sky plane.

\begin{figure}\epsscale{1}
\plotone{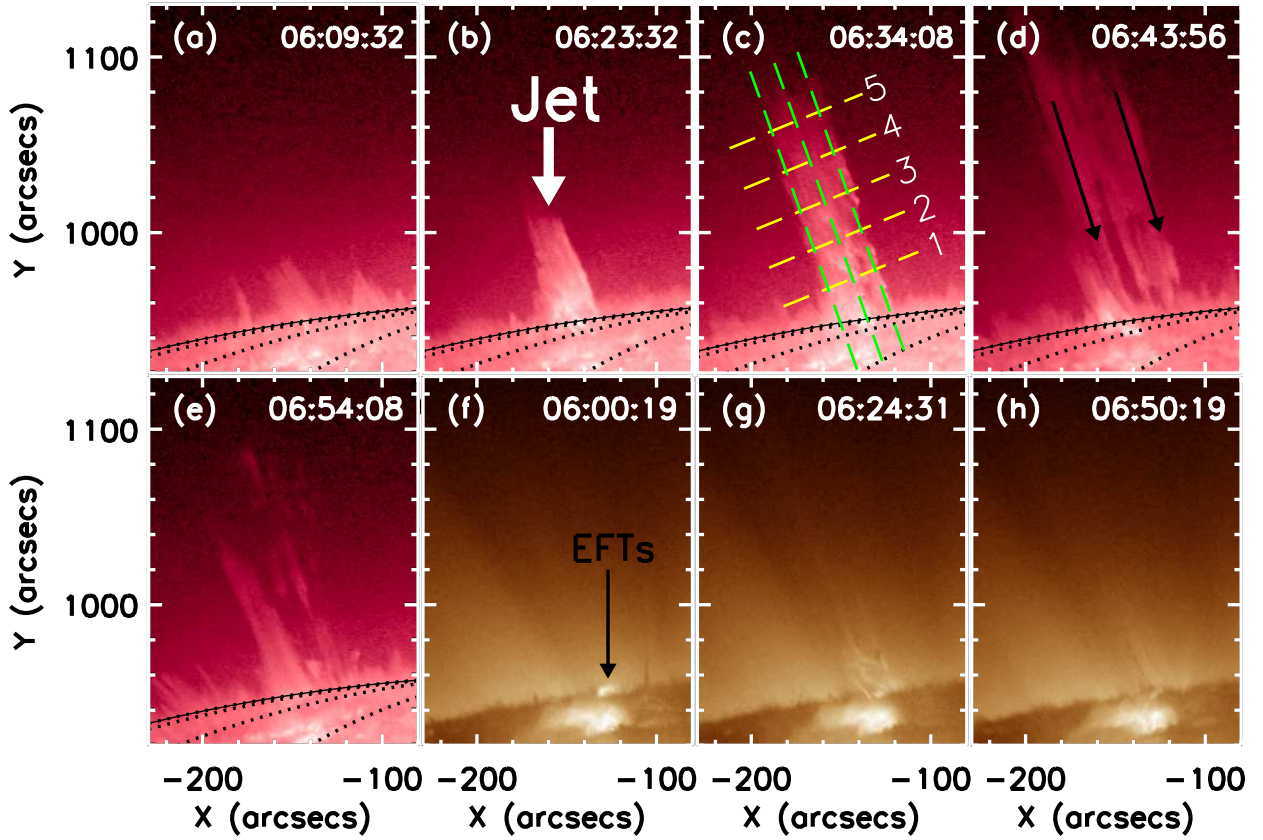}
\caption{Time sequences of {\sl SDO}/AIA 304 \AA\ ((a)--(e)) and 193 \AA\ ((f)--(h)) images. The green (yellow) dashed lines parallel (perpendicular) to the jet's main axis in frame (c) represent 2 (3) pixel cuts for the time--distance diagrams shown in \fig{vert} (\fig{tran}4). In addition, the heights of the cuts 1--5 from the jet base are 1.74, 3.48, 5.22, 6.97, and 8.71 $\times$ $10^{4}$ km respectively. The two parallel black arrows in frame (d) indicate the two paths of the falling plasmas, while the black arrow in frame (f) indicates the emerging flux tubes at 193 \AA\ wavelength. The FOV is $150\arcsec \times 210\arcsec$ for each frame.}
\label{304}
\end{figure}

\fig{304} shows the time evolution of the jet in 304 \AA\ and 193 \AA\ images. The jet first displayed as a cuspate structure (see the white arrow in \fig{304}), then it developed into a long cylindrical shape with many helical bright structures on the surface (see \fig{304}(c)). During this period, rotary motion around the jet's main axis and radial expansion of the jet at its western side are remarkable (refer to \fig{304} and the movie ``jet\_20100821.mpg'' in the online version of the journal). It is interesting to find that both the rising and the radial expansion periods of the jet stopped as the rotary motion drew to the end, which seems to imply that these coupled motions were caused by a single physical mechanism. At 06:43 UT, the jet reached the maximum height ($\sim$ 1.7 $\times 10^{5}$ km), then it began to fall back along two distinct straight paths without helical signal anymore (see the two parallel black arrows in \fig{304}(d)). We notice that a narrow cavity with a length of about 9 $\times 10^{4}$ km between the two paths was conspicuous in 304 \AA\ images around 06:43 UT but unidentifiable at other wavelengths. Since the 304 \AA\ line is not only out of Local Thermodynamic Equilibrium and optically thick but also sensitive to the material motion \citep{labr07,labr10}, therefore, the peculiar cavity in the 304 \AA\ images was possibly due to the longer integration length along the line of sight when observing the dynamic cylinder structure from the side. Based on this result, we propose that the mass of the jet should mainly distribute on the shell of a cylinder. In addition, a pre-existing bright point, which represents a small bipole with closed field, is observed near the footpoint of the jet in 193 \AA\ images (\fig{304} (f)). It became larger and brighter just before the start of the jet and then disappeared several minutes later, which suggests that the jet was the result from the interaction between the bipole and its surrounding background open field.

\begin{figure}\epsscale{1}
\plotone{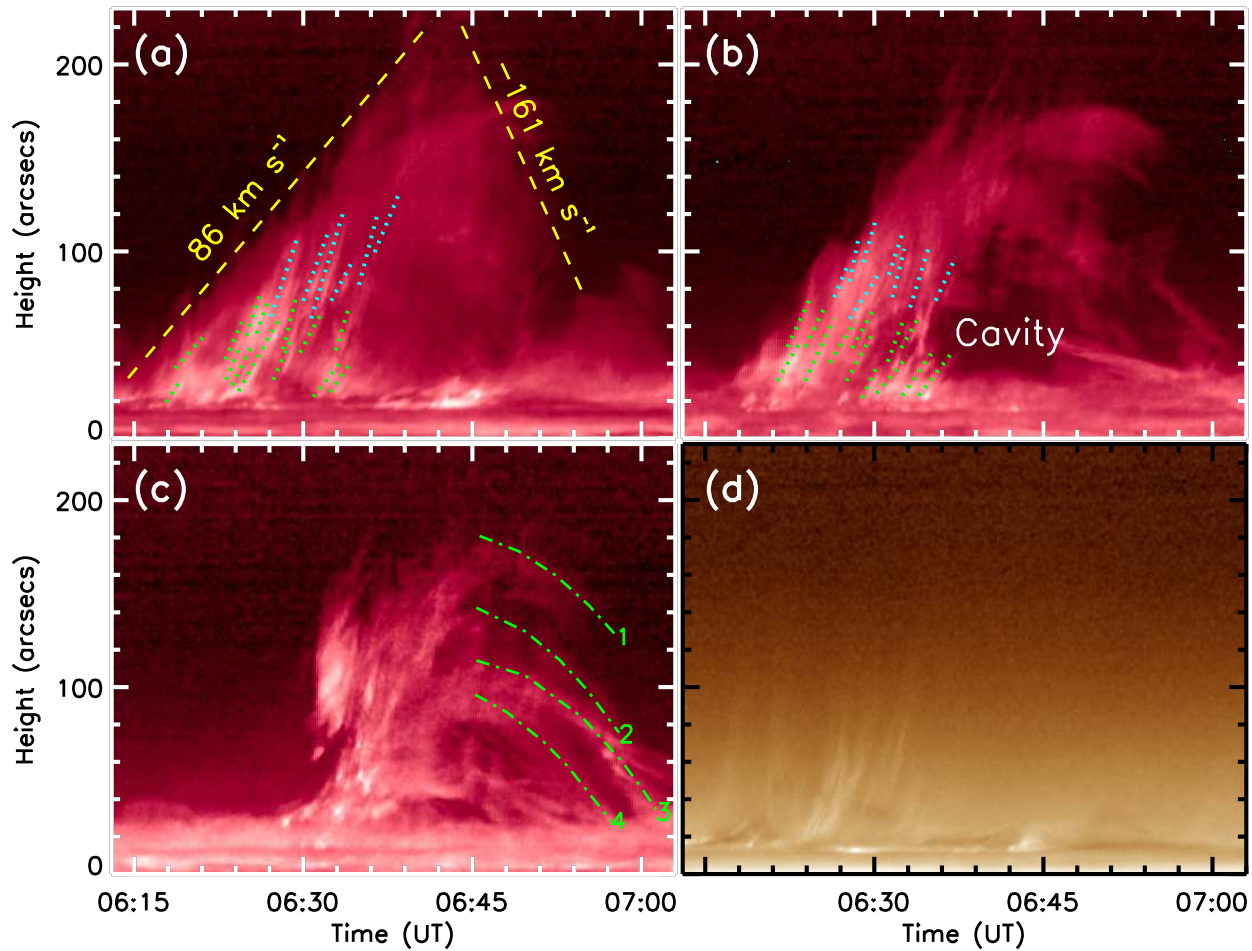}
\caption{Time--distance diagrams obtained from the cuts along the jet's main axis (\fig{304}(c)). Frames (a)--(c) correspond to the left, middle, and right cuts, respectively, at 304 \AA\ wavelength, while frame (d) is obtained from the middle cut at 193 \AA\ wavelength. The green and blue dotted lines are linear fit to the bright stripes, while the yellow dashed line is for the jet front. In addition, the green (blue) dotted lines indicate that the heights of the stripes from the jet base are below (above) $3.5 \times 10^4$ km. The dash-dotted curves in frame (c) are least-squares polynomial fit to the falling plasmas.}
\label{vert}
\end{figure}

\begin{figure}\epsscale{1}
\plotone{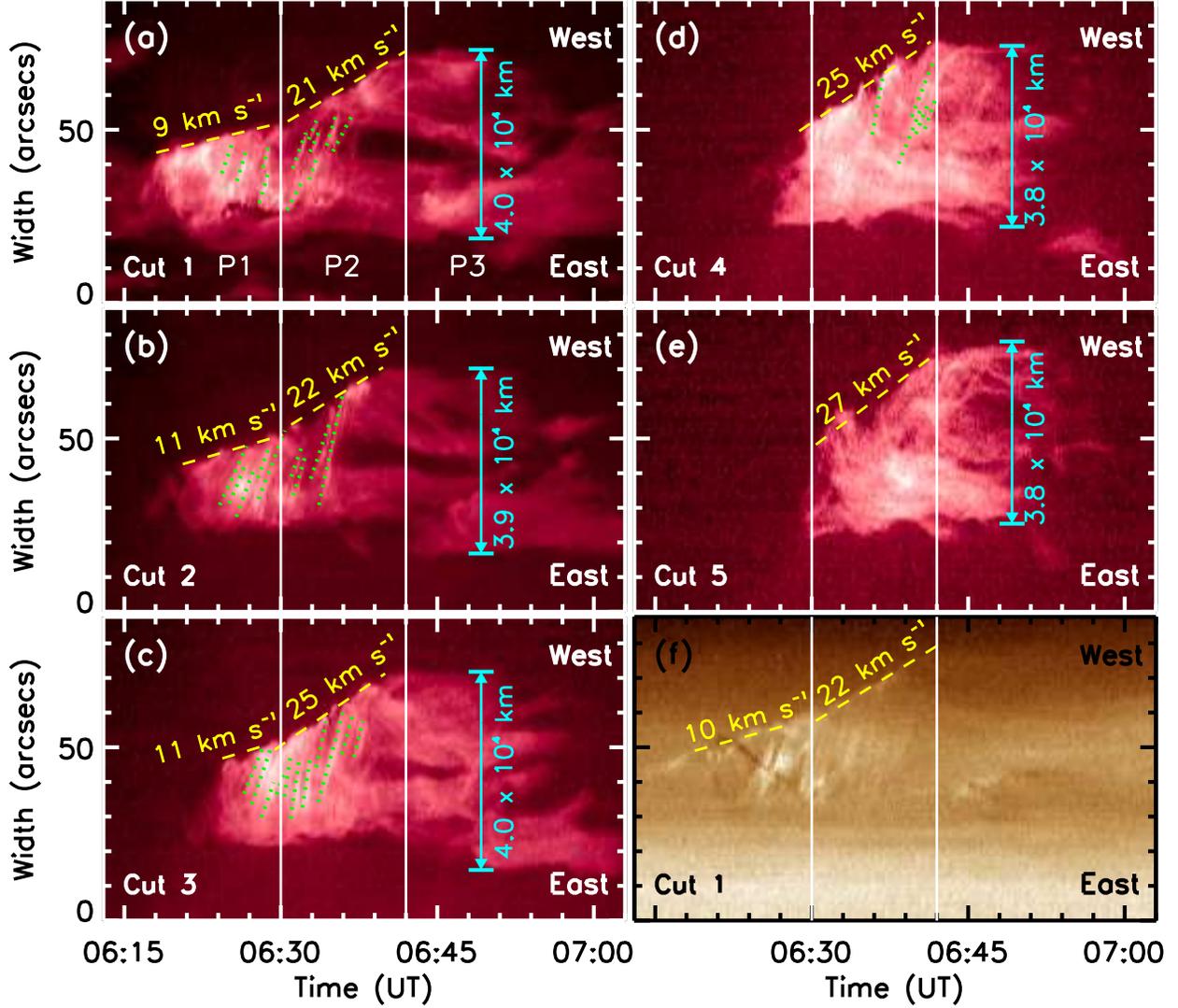}
\caption{Time--distance diagrams obtained from the cuts perpendicular to the jet's main axis (\fig{304} (c)). Frames (a)--(e) correspond to the cuts 1--5, respectively, at 304 \AA\ wavelength, while frame (f) is obtained from cut 1 at 193 \AA\ wavelength. The yellow dashed lines in each frame are the linear fit to the jet's western side, while the green dotted lines are for the bright stripes. The two white bars in each frame divide the radial expansion into three phases.}
\label{tran}
\end{figure}

The kinematics of the jet are displayed in the time--distance diagrams along (\fig{vert}) and across (\fig{tran}) the jet's main axis. In \fig{vert}, by applying a linear (least-squares polynomial) fit to the data, the rising (falling) speed of the jet front along the eastern cut is estimated to be \aspeed{86 (161)}, while the acceleration of the downward flows along the western cut is 0.026 km s$^{-2}$, which is far less than the solar gravitational acceleration ($\sim$0.27 km s$^{-2}$). Also, the cavity is obvious in 304 \AA\ time--distance diagram (\fig{vert}(b)) but undetectable in 193 \AA\ at the same position (\fig{vert}(d)). In \fig{tran}, an intriguing characteristic is the radial expansion of the jet along its western side, which underwent three distinct phases and each lasted for about 12 minutes. For the convenience, we call them the gradually expanding phase, the fast expanding phase, and the steady phase respectively. The gradually expanding phase can only be traced at low heights (cuts 1--3), while the fast expanding and steady phases are able to distinguish at all heights (cuts 1--5). The expanding speeds for the gradually and fast expanding phases are \speed{10 and 24}. We note that the transition from the gradually phase to the fast expanding phase was very quick, which is possibly due to a sudden acceleration of the magnetic reconnection between the closed field of the bipole and its ambient open field. During the steady phase, the jet kept its width to be about $4.0 \times 10^{4}$ km for 12 minutes before falling back to the solar surface. Furthermore, these phases are observable in other wave band unlike the cavity mentioned above (see \fig{tran}(f)), which indicate that they were indeed resulted from the mass motion rather than temperature change.

 When the highly twisted closed field lines of the bipole reconnected with the untwisted open field lines, nonlinear torsional Alfv$\rm \acute{e}$n waves propagating along the reconnected field lines would be produced. These waves can not only drive the jet but also compress locally the plasma in the jet, so the apparent speeds of the helical bright structures in the jet would be different from the actual plasma motion \citep{pari09}. However, to estimate the rotary speeds of these helical structures, we simply suppose that these helical structures are the results of the rotating plasmas. Consequently, the speeds can be obtained by tracing the bright stripes on the time--distance diagrams and making a linear fit to the data. The results are listed in Table \ref{tbl}. The axial (transverse) speed component ($v_{\rm ax}$ ($v_{\rm tr}$)) is obtained from time--distance diagrams along (\fig{vert}) and across (\fig{tran}) the jet's main axis, while the resultant (angular) speed ($v_{\rm re}$ ($\omega$)) is calculated based on the relation $v_{\rm re} = \sqrt{v_{\rm ax}^2 + v_{\rm tr}^2}$ ($\omega = \frac{v_{\rm tr}}{r}$, here $r$ is the radius of the jet). The average axial and transverse speeds and the resultant speed of the rotating plasmas are \speed{171, 123, and 211} respectively. When considering the speeds in different height ranges and different expanding phases, we find that the values above 3.5 $\times$ 10$^{4}$ km from the jet base are obviously higher than those below this height. Therefore, the rotating plasmas were accelerated during the rising period, and the progressing pinch front, the {\bf J $\times$ B} force \citep{shib86}, and the torsional Alfv$\rm \acute{e}$n waves \citep{pari09} in the unwinding helices are considered to be the possible drivers. The mean angular speeds ($\omega$) of the rotating plasmas are $11.1 \times 10^{-3}$ rad s$^{-1}$ (period = 564 s), and there is little difference between the values for the gradually and fast expanding phases. Furthermore, we estimate the twist shed into the outer corona in the eruption. The upper boundary value of the twist stored could be obtained by multiplying $\omega$ by the unwinding time of the jet ($\sim$ 24 minutes), while the lower boundary is estimated by multiplying $\omega$ by the mean untwisting time of some individual helical structures ($\sim$ 11 minutes), since the reconnection occurs sequentially over different field lines and the new reconnected open lines will untwist orderly at a different moment. So the twist stored is between 1.17 and 2.55 turns, or 2.34--5.1$\pi$. Since magnetic helicity is conserved under reconnection, the released twist should equal to the twist stored in the twisted closed field of the bipole. In addition, by dividing the turns of the released twist by the jet's maximum length, it yields the force-free factor ($\alpha$) of the jet's magnetic field, which is between 0.7--1.5 $\times 10^{-8} $ m$^{-1}$, in agreement with a typical $\alpha$ value of active regions \citep{pevt03}.

It has been widely accepted that the energy source of solar eruptions is mostly transferred from magnetic free energy storing in non-potential magnetic field systems and releasing through magnetic reconnection. To investigate the energy relationship between the jet and its magnetic field, we calculate the jet's total energy, and compare it with the free energy stored in the jet's non-potential magnetic field. With the formula $\Phi(R) = \frac{LB_\phi(R)}{RB_z(R)}$, here $\Phi(R)$, $L$, and $R$ are the twist, length, and radius of the jet, $B_\phi$ and $B_z$ are the azimuthal and the longitudinal components of the jet's magnetic field \citep{prie84}, the azimuthal component ($B_\phi$) can be obtained by setting the local vertical components ($B_z$) as 5 G \citep{smit95}. Thus $B_\phi$ is between 3.8--8.3 G. Moreover, assuming the jet's shape to be a cylinder, the free energy storing in the non-potential magnetic field of the jet is estimated to be 1.0--4.5 $\times 10^{30}$ erg. On the other hand, the total energy (including the kinetic energy and the gravitational potential energy of the jet, and the thermal energy of a microflare associated with the jet) of the jet is about $2.0 \times 10^{29}$ erg if the electron density is set as $4 \times 10^{9}$ cm$^{-3}$ and the thermal energy of the associated microflare is $10^{28}$ erg \citep{shim00}. The results indicate that the free energy in the jet's non-potential magnetic field is one order of magnitude higher than the total energy of the jet. That is to say, the non-potential magnetic field of the jet can be the energy source for the jet.

\section{CONCLUSIONS AND DISCUSSIONS}
With the high temporal and spatial observations, we study the kinematics and fine structure of an unwinding polar jet, which can be interpreted by using the reconnection models invoking magnetic untwisting as the driving mechanism \citep{shib86,pari09,pari10}. The main results are summarized as follows. (1) The jet showed separate dense points along its body in H$\alpha$ observations. It represents the appearance of helical motions in the jet. (2) The shape of the jet resembled a cylinder with helical structures on the surface. Since the cavity can only be observed in 304 \AA\ line, we propose that the mass of the jet should mainly distribute on a cylindrical shell. (3) The erupting plasmas fell back to the solar surface along two distinct paths. The acceleration of the downward flows along the western path was far less than the solar gravitational acceleration, which is possibly resulted from the projection and the damping effects of the background magnetized plasmas. (4) The jet's radial expansion mainly occurred at its western side and underwent three distinct phases: the gradually expanding phase, the fast expanding phase, and the steady phase. Each phase lasted for about 12 minutes. (5) The rising, radial expansion, and unwinding motions of the jet are identified to stop approximately at the same time. We conjecture that these coupled motions are driven by a single physical mechanism. (6) The mean angular speed of the rotating plasmas around the jet's main axis is about $11.1 \times 10^{-3}$ rad s$^{-1}$ (period = 564 s), while the twist shed into the outer corona in the eruption is estimated to be 1.17--2.55 turns (or 2.34--5.1$\pi$). In addition, the $\alpha$ of the jet's magnetic field (0.7--1.5 $\times 10^{-8} $ m$^{-1}$) is consistent with the $\alpha$ of active regions. (7) Based on the observational results, we calculate the jet's total energy and the free energy stored in the non-potential magnetic field of the jet. It suggests that the non-potential magnetic field of the jet is enough to supply the total energy for the jet.

In terms of theory, a twisted, force-free, cylindrical flux tube will become unstable to kink instability if the twist of the tube reaches up to 1.3 turns, or $2.6\pi$ \citep{prie84}. In the numerical simulation presented by \cite{pari09}, the authors found that the threshold for producing a coronal jet is 1.4 turns (2.8$\pi$) of twist. In our case, the twist of the closed field not only agrees with the theoretical result, but also confirms that it indeed affects the tube's stability. It seems incompatible that the falling speed of the jet at the eastern side is higher than the rising speed. This is possibly that the heated erupting plasma has cooled down and therefore diminished the emission from the 304 \AA\ images during the falling period \citep{labr07}. Since the jet is driven by magnetic twist through reconnection process, we propose that the rising, radial expanding, and unwinding of the jet are the different manifestations of the reconnection. In addition, the expanding phases are thought to be closely associated with the reconnection between the closed field of the bipole and its ambient open field, and the transition from the gradually expanding phase to the fast expanding phase is possibly owning to a sudden acceleration of the reconnection process. In the energy calculation, the jet's shape is considered to be a cylinder. However, our observation indicates that the mass of the jet is mainly distributed on a cylindrical shell (viz., a hollow cylinder). If the thickness of the shell is assumed to be half of the jet's radius, the free energy stored in the jet's non-potential magnetic field is estimated to be 0.7 -- 3.4 $\times 10^{30}$ erg, while the jet's total energy is about $1.7 \times 10^{29}$ erg. This result is consistent with the conclusion summarized above. To fully understand the jet phenomenon, more observational and theoretical investigations with high temporal and spatial observations are needed in the future work.

\acknowledgments
The AIA data are courtesy of {\sl SDO} (NASA) and the AIA consortium, and we thank the SMART team for the H$\alpha$ data support. We thank the referee for many helpful comments and valuable suggestions. This work is supported by the NSFC under Grants 10933003, 11078004, and 11073050, MOST (2011CB811400), and Open Research Program of Key Laboratory of Solar Activity of National Astronomical Observatories of Chinese Academy of Sciences (KLSA2011\_14).

\begin{table}
\begin{center}
\caption{List of Various Speeds and Other Physical Parameters of the Jet\label{tbl}}
\tabcolsep 1mm
\newcommand{\tabincell}[2]{\begin{tabular}{@{}#1@{}}#2\end{tabular}}
\begin{tabular}{ccccccccc}
\tableline
\tableline
 Items  & \tabincell{c}{$v_{\rm ax}$\\(km s$^{-1}$)} & \tabincell{c}{$v_{\rm tr}$\\(km s$^{-1}$)} & \tabincell{c}{$v_{\rm re}$\\(km s$^{-1}$)}  & \tabincell{c}{$v_{\rm ex}$\\(km s$^{-1}$)} & \tabincell{c}{$r$\\(10$^4$ km)} & \tabincell{c}{$\omega$\\(10$^{-3}$ rad s$^{-1}$)} & \tabincell{c}{$T$\\(s)}\\
\tableline
Below $3.5 \times 10^4$ km  &146 &104 &179 &$\cdot\cdot\cdot$ &1.41 &8.3  &760  \\
Above $3.5 \times 10^4$ km  &218 &143 &261 &$\cdot\cdot\cdot$ &1.13 &14.7  &426 \\
\tabincell{c}{The gradually\\expanding phase} &169 &118 &206 &10  &1.03 &10.6  &593 \\
\tabincell{c}{The fast\\expanding phase}  &173 &127 &215 &24  &1.41 &11.4  &550 \\
Average value                    &171 &123 &211 &17  &1.28 &11.1  &564 \\
\tableline
\end{tabular}

\tablecomments{In this table, $v_{\rm ax}$, $v_{\rm tr}$, $v_{\rm re}$, and $\omega$ are the axial component, transverse component, resultant, and angular speeds of the rotating plasmas, while $v_{\rm ex}$, $r$, and $T$ are the jet's radial expanding speed, radius, and period of rotation, respectively.}
\end{center}
\end{table}

\end{document}